\documentclass[pra,twocolumn,a4paper,groupedaddress,showpacs,floatfix]{revtex4}
\usepackage{graphicx,amsmath,amssymb,amsfonts,dsfont,subfigure}

\newcommand{\ket}[1]{\ensuremath{|#1\rangle}}

\newcommand{\mean}[1]{\ensuremath{ \langle #1  \rangle}}

\usepackage{color}

\begin{document}

\title{Beyond mean-field bistability in driven-dissipative lattices: bunching-antibunching transition and quantum simulation}

\author{J. J. Mendoza-Arenas$^{1,4}$}
\author{S. R. Clark$^{1,3}$}
\author{S. Felicetti$^{5}$}
\author{G. Romero$^{6}$}
\author{E. Solano$^{5,7}$}
\author{D. G. Angelakis$^{2,8}$}
\author{D. Jaksch${}^{1,2}$}

\affiliation{${}^1$Clarendon Laboratory, University of Oxford, Parks Road, Oxford OX1 3PU, United Kingdom}
\affiliation{${}^2$Centre for Quantum Technologies, National University of Singapore, 3 Science Drive 2, Singapore 117543}
\affiliation{${}^3$Department of Physics, University of Bath, Claverton Down, Bath, BA2 7AY, United Kingdom}
\affiliation{${}^4$Departamento de F\'{i}sica, Universidad de los Andes, A.A. 4976, Bogot\'a D. C., Colombia}
\affiliation{${}^5$Department of Physical Chemistry, University of the Basque Country UPV/EHU, Apartado 644, 48080 Bilbao, Spain}
\affiliation{${}^6$Departamento de F\'isica, Universidad de Santiago de Chile (USACH), Avenida Ecuador 3493, 9170124, Santiago, Chile}
\affiliation{${}^7$IKERBASQUE, Basque Foundation for Science, Maria Diaz de Haro 3, 48013 Bilbao, Spain}
\affiliation{${}^8$School of Electrical and Computer Engineering, Technical University of Crete, Chania, Crete, Greece, 73100}


\date{\today}

\begin{abstract}
In the present work we investigate the existence of multiple nonequilibrium steady states in a coherently-driven $XY$ lattice of dissipative two-level systems. A commonly-used mean-field ansatz, in which spatial correlations are neglected, predicts a bistable behavior with a sharp shift between low- and high-density states. In contrast one-dimensional matrix product methods reveal these effects to be artifacts of the mean-field approach, with both disappearing once correlations are taken fully into account. Instead a bunching-antibunching transition emerges. This indicates that alternative approaches should be considered for higher spatial dimensions, where classical simulations are currently infeasible. Thus we propose a circuit QED quantum simulator implementable with current technology, to enable an experimental investigation of the model considered.   
\end{abstract}

\maketitle

\section{Introduction}
Nonequilibrium steady states (NESS) of driven dissipative many-body quantum systems are of increasing interest, both theoretically and experimentally, due to their potentially strong response to external changes in technologically relevant contexts. For example, intense research has been recently performed on quantum transport through nanoscale systems driven at their boundaries.  For molecular or quantum dot junctions, predictions of correlation-induced current oscillations and current-voltage bistability have been made~\cite{kurth2010, wilner2013, khosravi2012}, while for spin chains sharp changes in magnetic conductance are expected due to a nonequilibrium phase transition \cite{benenti2009negative, vznidarivc2011spin, prosen_zni2009, mendoza2013, mendoza2013jstat}. Other interesting examples include recent studies of remnants of equilibrium phase transitions in dissipative settings \cite{ssn2010signatures,tomadin2010prl,liu2011quantum,tomadin2011nonequilibrium,grujic2012,joshi2013}, repulsively induced photon super-bunching \cite{grujic2013}, and potential super-solid phases in driven resonator arrays \cite{jin2013photon}. A rigorous study of these open nonequilibrium quantum systems is very challenging, particularly in high dimensions, given the exponential growth of the associated Hilbert space which prohibits direct classical simulation. So a natural question is whether their physics can be correctly determined from approximate schemes with reasonable computational cost.  

A frequently used method of describing interacting quantum lattice systems is to employ a mean-field product ansatz, in which spatial correlations are neglected. In equilibrium this approach can often yield qualitatively correct features such as the presence of phase transitions, although it can fail to correctly identify the exact location and critical exponents, especially in low dimensions~\cite{LeBellac04,Sachdev11}. For nonequilibrium systems the situation is quite different. Even though mean-field calculations offer a first step to help uncover the intricate dynamics taking place, and have been used in several recent studies of driven-dissipative models~\cite{ssn2010signatures,tomadin2010prl,tomadin2011nonequilibrium,liu2011quantum,Dombi13,lee2011,boite2012steady,nissen2012nonequilibrium,carr2013,Fazio2014,LeBoite2014_PRA,marcuzzi2014,Naether2015,everest2015}, it is not clear that they can provide even a qualitatively correct physical description. Furthermore, reasoning based on the Ginzburg criterion, according to which equilibrium critical phenomena are correctly described by mean-field theory above a critical spatial dimension~\cite{LeBellac04}, cannot be relied upon in nonequilibrium settings.
This has motivated the recent development of several new methods to analyze driven-dissipative models, namely the self-consistent Mori projector technique~\cite{Degenfeld2014}, a variational minimization calculation~\cite{Weimer2015_1,Weimer2015_2,Weimer2015_3}, a corner-space renormalization~\cite{finazzi2015}, and algorithms based on Matrix Product Operators (MPO) for one-dimensional lattices~\cite{Mascarenhas2015,Cui2015}.

A notable effect predicted by mean-field descriptions of driven-dissipative models is that of bistability~\cite{Dombi13,lee2011,boite2012steady,nissen2012nonequilibrium,carr2013,Fazio2014,LeBoite2014_PRA,marcuzzi2014}.
Here the existence of two distinct NESS is observed in a particular parameter regime, with the actual state obtained depending on the history of the system. However it is usually expected for systems described by a Lindblad master equation~\cite{breuer2002,gardiner04}, such as those featuring mean-field bistability~\cite{lee2011,boite2012steady,nissen2012nonequilibrium,carr2013,Fazio2014,LeBoite2014_PRA,marcuzzi2014}, to have a unique NESS~\cite{Degenfeld2014,Spohn1977}. This raises the question of whether this bistability is physical or an artifact of the mean-field approximation, originating from the effective non-linearity introduced by self-consistently factorizing the long-range correlations. In fact, recent studies have found that the bistability is washed out when correlations are taken into account~\cite{Degenfeld2014,Weimer2015_1,Weimer2015_2,maghrebi2015}. The impact of long-range correlations, however, is still not clear~\cite{Bonnes2014}, and interesting effects that might emerge are yet to be uncovered. 

Since bistability is a basic and ubiquitous feature of driven non-linear systems~\cite{glendinning1994stability}, in this paper we examine in detail whether multiple NESS exist in a dissipative coherently-driven quantum lattice system. In particular, we contrast predictions from mean-field analysis with results from tensor network theory (TNT) methods in one spatial dimension~\cite{zwolak2004mixed,verstraete04}, in which states are described by a matrix product ansatz. These calculations show that as long-range correlations are progressively handled more exactly, a single NESS emerges. In its place a bunching-antibunching transition is found. This points to major qualitative errors in describing driven dissipative interacting systems if these correlations are neglected. Also, given the formidable challenge to classically simulate such open quantum lattices, even using sophisticated TNT methods~\cite{vidal2003efficient,vidal2004tebd,tnt,verstraete08,cirac09,schollwock11,evenbly11,orus13,biamonte11}, obtaining sound insight into the physics of these systems without uncontrolled approximations necessitates experimental realization and verification. Thus we also discuss how a quantum simulation of the model discussed here in higher spatial dimensions could be implemented using current circuit QED technology. This would also allow experimentalists to confirm our predictions for 1D systems.

The paper is organized as follows. In Section~\ref{model_section} we describe the driven-dissipative model to be considered. In Section~\ref{mf_section} we show the bistable behavior resulting from mean-field calculations of the NESS. In Section~\ref{mp_section} we discuss the impact of correlations in one-dimensional lattices, namely the breaking of the bistability and the emergence of a bunching-antibunching transition. In Section~\ref{simulation_section} we describe a possible experimental implementation based on transmon qubits in circuit QED. Finally, we present our conclusions in Section~\ref{conclu}.

\section{Driven-dissipative lattice model} \label{model_section}

\begin{figure}[t]
\centering
\includegraphics[width=0.35\textwidth]{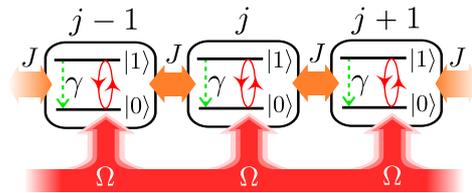}
\caption{A schematic of the generalized $XY$ model showing three adjacent sites of a one-dimensional chain. Each site $j$ contains a TLS which is coherently coupled to its $z$ ($=2$ for 1D) nearest-neighbours with amplitude $J$. Circular red arrows represent coherent driving $\Omega$, while the dashed vertical lines depict the dissipation $\gamma$.}
\label{fig1}
\end{figure}

In this work we restrict our attention to a concrete minimal model which nevertheless possesses features common to more complex quantum lattice models. Specifically we consider a lattice of two level systems (TLS), each one with upper level $\ket{1}$ and lower level $\ket{0}$, featuring coherent hopping linking adjacent sites, bulk coherent driving, and incoherent loss processes. The model is schematically shown in Fig.~\ref{fig1}.

The TLS are described by Pauli transition matrices $\sigma_j^\pm$ and an external driving field detuned by $\Delta$ from the TLS resonance. In a frame rotating with the driving field the Hamiltonian is ($\hbar = 1$)
\begin{eqnarray} \label{hami_first}
{\mathcal{H}} & = & \sum_{j} \left [ \Delta\sigma^+_j \hat{\sigma}^-_j + \Omega (\sigma^+_j + \sigma^-_j )\right ] - J \sum_{\langle j, j'\rangle } \sigma_j^+ \sigma_{j'}^-.
\label{eq:toy_model_ham}
\end{eqnarray}
Here $J$ is the coherent tunneling amplitude between neighboring TLS, and $\Omega$ is the Rabi frequency of the driving field. The index $j$ runs over the discrete lattice sites, and $\langle j, j'\rangle$ denotes the set of nearest neighbors. Since the hopping term can be rewritten as a coupling of $XY$ type, this is known as the $XY$ Hamiltonian.  

Finally, we incorporate a generic local loss term $\gamma$ which acts to incoherently de-excite the upper level $\ket{1}$ of each TLS to its lower level $\ket{0}$. The evolution of the total system density matrix $\rho$ is then described by a quantum master equation $\dot{\rho} = \mathcal{L} [\rho]$ in Lindblad form, where
\begin{equation}
\mathcal{L}[\rho] = \frac{1}{i} [\mathcal{H}, \rho] + \frac{\gamma}{2} \sum_{j} \left ( 2\sigma_j^- \rho\sigma_j^+ - \sigma_j^+ \sigma_j^- \rho - \rho \sigma_j^+ \sigma_j^- \right ).
\label{eq:master_equation}
\end{equation}
A NESS $\rho_{\rm NESS}$ of the system satisfies $\mathcal{L}[\rho_{\rm NESS}] = 0$, and all observables $O$ are measured with respect to this state, so $\mean{O} \equiv {\rm Tr} (O \rho_{\rm NESS})$. We also note that our calculations are performed with open boundary conditions. Thus the system does not satisfy translational invariance, but is symmetric with respect to its center~\footnote{This choice is made considering that the use of periodic boundary conditions would make the TNT simulations more challenging~\cite{schollwock11}.}. In addition, we have verified that our results remain essentially unchanged when considering larger system sizes than those used in the calculations discussed in the manuscript. 

\section{Mean-field approach} \label{mf_section}

We start by discussing the physics of the driven-dissipative system resulting from a single-site mean-field analysis. This can be done by means of two different methods, namely by obtaining the NESS of the system through a simulation of the mean-field master equation, or by performing a Monte Carlo wave function calculation~\cite{plenio98,breuer2002,gardiner04,Daley2014traj} in the mean-field approximation. We now show that both methods indicate the existence of bistable behavior, for one- and two-dimensional lattices. 

\subsection{Product density matrix solution} \label{master_eq_mf_section}

Initially we assume that for every time $t$, the density matrix of a driven-dissipative $XY$ lattice of $N$ sites can be factorized in the form
\begin{equation} \label{mf_ansatz}
\rho(t)=\bigotimes_{j=1}^N\rho_j(t).
\end{equation}
This mean-field ansatz captures the local physics, but neglects all classical and quantum intersite correlations. When inserting this approximation into the master equation~\eqref{eq:master_equation}, as described in Appendix~\ref{mfa_me}, we obtain that the coherent dynamics of each TLS is governed by an effective mean-field (mf) local Hamiltonian. For site $j$ this is given by
\begin{align} \label{mean_field_hamiltonian_main}
\begin{split}
\mathcal{H}_j^{\text{mf}}=\Delta\sigma_j^+\sigma_j^-&+\Omega_j\sigma_j^++\Omega_{j}^{*}\sigma_j^-,
\end{split}
\end{align}
where the nearest-neighbor hopping is effectively taken into account by modified site-dependent coherent driving amplitudes
\begin{equation} \label{effective_driving_mf}
\Omega_j=\Omega-J\sum_{j'}\langle\sigma_{j'}^{-}\rangle,
\end{equation} 
with the sums performed over the sites $j'$ coupled to site $j$. Thus the equation of motion of each TLS (see Eqs.~\eqref{local_mf_eq} and~\eqref{mf_local_evolution_obc}) becomes dependent on expectation values of neighboring sites, leading to nonlinear dynamics. The corresponding NESS is obtained by performing the time evolution for a particular initial state $\rho(0)$ in the long-time limit, until convergence is reached.

\begin{figure}
  \centering
\includegraphics[scale=1.05]{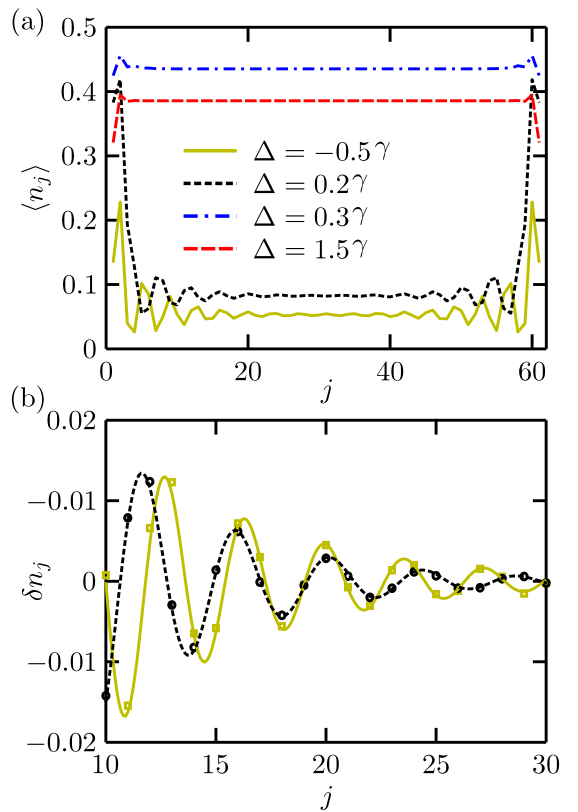}
\caption{(a) Density profiles of the one-dimensional driven-dissipative model, for different values of $\Delta/\gamma$ and the R-L sweep. The results correspond to $J/\gamma=2$, $\Omega/\gamma=1$, and $N=61$.
(b) Exponential decay of oscillations in low-density regime. The colors and line types correspond to the same parameters as in (a). The symbols correspond to the results of the simulations, and the lines to the fits to Eq.~\eqref{decay_oscil}. For $\Delta/\gamma=-0.5$, $A=0.08(2)$, $\phi=-7.9(2)$, $r=7.0(8)$ and $k=1.73(2)$. For $\Delta/\gamma=0.2$, $A=-0.11(2)$, $\phi=-12.6(2)$, $r=5.6(3)$ and $k=1.48(1)$.}
\label{fig2}
\end{figure}

It is usually expected that in the absence of very particular symmetries~\cite{Buca2012}, an open system governed by a Lindblad master equation such as Eq.~\eqref{eq:master_equation} relaxes to a unique NESS, independent of the initial state~\cite{Spohn1977,prosen_zni2009,Degenfeld2014}. However, as reported in previous mean-field studies of driven-dissipative models~\cite{lee2011,boite2012steady,nissen2012nonequilibrium,carr2013,Fazio2014,LeBoite2014_PRA,marcuzzi2014}, a bistable behavior emerges in certain parameter regimes, corresponding to the existence of two different stable NESSs. Whether the systems relaxes to one or the other NESS depends on which domain of attraction the initial condition lies in.

To verify whether the system under consideration features a bistable behavior, we proceed as follows. First, for a one-dimensional lattice with fixed values of $J$, $\Omega$ and $\gamma$, we take a detuning value $\Delta_0$ such that $\Delta_0/\gamma\gg1$, and obtain its NESS for different random initial states. After verifying that such a NESS is unique, we use it as the initial state for the calculations of detuning values $\Delta<\Delta_0$, sweeping from higher to lower values of $\Delta$; this is the right to left (R-L) sweep. Subsequently we perform a similar sweep process but in the opposite direction. Thus we take a new detuning $\Delta_0/\gamma<0$ whose unique NESS serves as the initial state for simulations of values of $\Delta>\Delta_0$; this is the left to right (L-R) sweep. A bistable zone is manifested as a parameter regime where the solutions of the two sweeps are different, i.e. a hysteresis region.

\begin{figure}
  \centering
\includegraphics[scale=1.07]{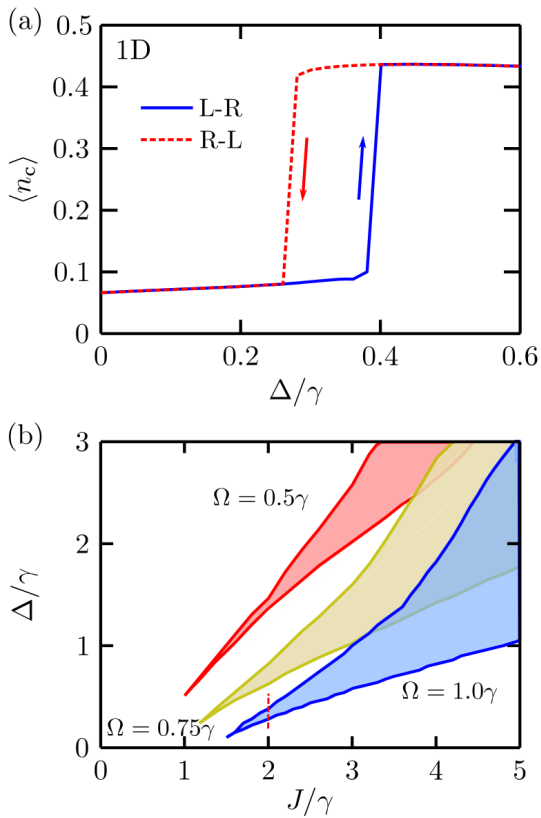}
\caption{(a) Hysteresis of the central-site density $\langle n_{\text{c}}\rangle$ as a function of $\Delta/\gamma$, for the one-dimensional lattice. The results correspond to $J/\gamma=2$, $\Omega/\gamma=1$, and $N=61$. (b) Bistable regions in the $(\Delta/\gamma,J/\gamma)$ plane for different values of $\Omega$. The vertical dashed line indicates the bistable region depicted in panel (a).}
\label{figure3}
\end{figure}

We first discuss the results for the R-L sweep. In Fig.~\ref{fig2}(a) we show the corresponding local densities $\langle n_j\rangle=\langle\sigma_j^+\sigma_j^-\rangle$ for all sites $j$ and different values of $\Delta/\gamma$. Here we can already observe two qualitatively different types of NESS. For $\Delta/\gamma>(\Delta/\gamma)_c\approx0.28$, the state corresponds to a flat high-density configuration. On the other hand, for $\Delta/\gamma<(\Delta/\gamma)_c$, the bulk of the lattice is in a low-density state, which as depicted in Fig.~\ref{fig2}(b) shows density oscillations $\delta n_j$ that decay exponentially towards the bulk density average $\bar{n}$, in the form
\begin{equation} \label{decay_oscil}
\delta n_j\equiv n_j-\bar{n}=Ae^{-j/r}\sin(kj+ \phi).
\end{equation}

Notably, as shown in Fig.~\ref{figure3}(a) for the central site (i.e. for site $j=\lceil N/2\rceil$, with density $\langle n_{\text{c}}\rangle$), the shift from the low- to the high-density NESS taking place at the critical value $(\Delta/\gamma)_c$ is very sharp.

\begin{figure}
  \centering
\includegraphics[scale=1.05]{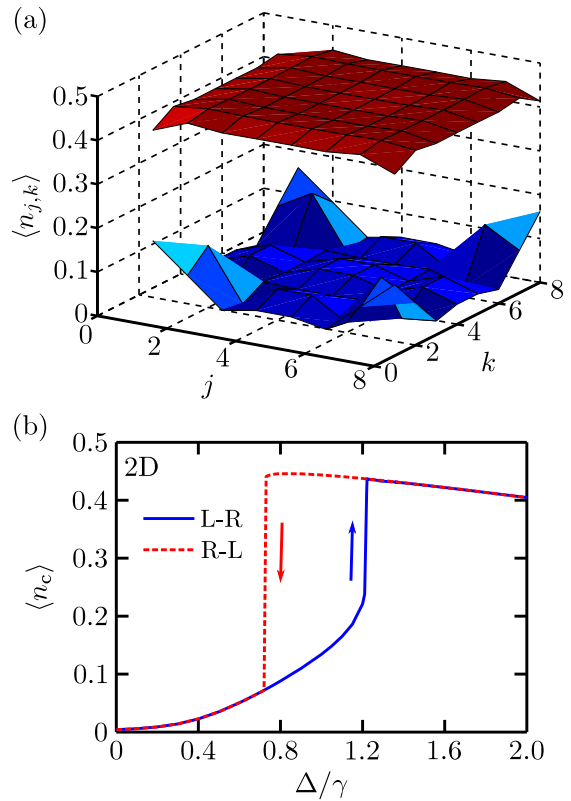}
\caption{(a) Density profiles of the two-dimensional driven-dissipative model, for $J/\gamma=2$, $\Omega/\gamma=1$, a $8\times8$ lattice ($N=64$) and the R-L sweep. The low-density profile corresponds to $\Delta/\gamma=0.7$, and the high-density regime to $\Delta/\gamma=0.8$. (b) Density $\langle n_{\text{c}}\rangle$ for a central site of the lattice, the L-R and R-L sweeps and the same parameters as in (a), as a function of $\Delta/\gamma$, indicating hysteresis. For the R-L sweep the critical detuning separating the low- and high-density regimes is $(\Delta/\gamma)_c=0.73$, while for the L-R sweep it is $(\Delta/\gamma)_c=1.22$.}
\label{figure4}
\end{figure}

For the L-R sweep similar results are obtained, with an important difference. Namely, as shown in Fig.~\ref{figure3}(a), a different critical value for the sharp shift between the low- and high-density regimes is found, $(\Delta/\gamma)_c\approx0.40$. Thus the mean-field treatment of the driven-dissipative model suggests the existence of bistable behavior. In Fig.~\ref{figure3}(b) we depict the bistability zones for a wider parameter regime, i.e. for different driving amplitudes $\Omega$ in the $(\Delta/\gamma, J/\gamma)$ plane. These zones have triangle-like shapes, and become broader and shift to lower $\Delta/\gamma$ for larger values of $\Omega$.  

Similar physics is obtained for two-dimensional lattices. As shown in Fig.~\ref{figure4}(a), the density profiles $\langle n_{j,k}\rangle=\langle\sigma_{j,k}^+\sigma_{j,k}^-\rangle$ for all pairs of sites $(j,k)$ indicate the existence of two distinct nonequilibrium phases of the driven-dissipative model: a high-density NESS with flat profile, and a low-density NESS with decaying oscillations towards the center of the lattice. In addition, as depicted in Fig.~\ref{figure4}(b), the location of the sharp shift between both types of states depends on the direction of the parameter sweep, indicating bistable behavior.

\subsection{Monte Carlo wave function approach} \label{trajectories_section}

An alternative way to study an open quantum system described by a Lindblad master equation of the form~\eqref{eq:master_equation} corresponds to a Monte Carlo-type calculation. Here instead of time-evolving the density operator of the lattice, the evolution of several independent realizations (or trajectories) of the system is performed, each described by a pure state. Due to the dissipative processes from the environmental coupling, the evolution in each trajectory is governed by a modified Hamiltonian, and at random times quantum jumps describing such a coupling are applied to the lattice. Finally, expectation values are obtained by performing averages over the sample of simulated trajectories. This technique is well known in the quantum optics community, and is described in detail in several references, e.g. see~\cite{plenio98,breuer2002,gardiner04,Daley2014traj,Bonnes2014_arxiv}.

To perform a mean-field Monte Carlo wave function calculation, we simply assume that at every time the pure state of each realization is a product. Namely, for trajectory $r$ the state for a lattice of $N$ sites is 
\begin{equation} \label{mf_pure_state}
|\Psi^{(r)}(t)\rangle=|\psi^{(r)}_1(t)\rangle\otimes|\psi^{(r)}_2(t)\rangle\otimes\cdots\otimes|\psi^{(r)}_N(t)\rangle.
\end{equation}
First we take a random product of the latter form as the initial state of each trajectory. Then we perform the time evolution as described in Appendix~\ref{mfa_tm}, for long-enough times to obtain the NESS of the system. In our particular case, we evolved for a total time of $T=200/\gamma$, with a time step $\delta t=2\times10^{-3}$. Finally we obtain the NESS expectation values of interest by performing an average over a sample of $N_{\text{traj}}=1000$ trajectories. To further smoothen the results, we also average over the final $30\%$ time steps. 

\begin{figure}
\centering
\includegraphics[scale=1.12]{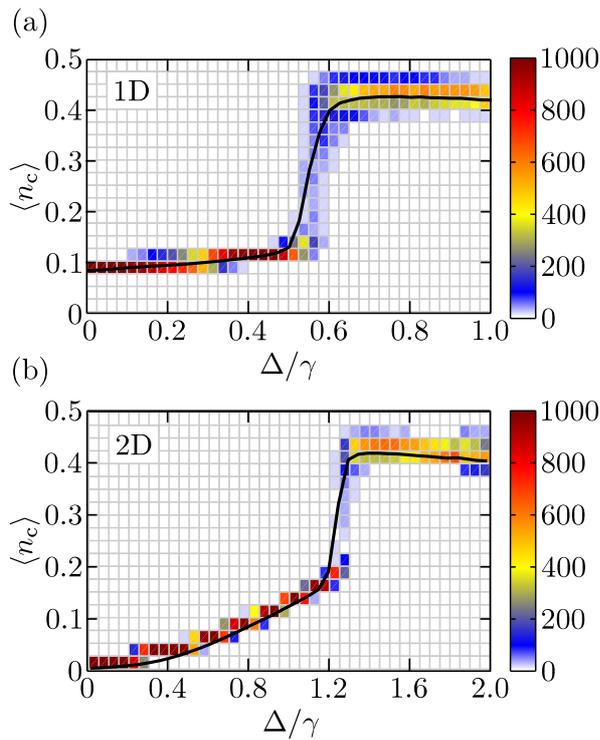}
\caption{Distribution of central populations $\langle n_{\text{c}}\rangle$ obtained from each time-averaged trajectory, as a function of $\Delta/\gamma$, for $J/\gamma=2$ and $\Omega/\gamma=1$, and average over $N_{\text{traj}}=1000$ trajectories (black solid lines). (a) Results for a one-dimensional lattice of $N=61$ sites. (b) Results for a two-dimensional $8\times8$ lattice of $N=64$.}
\label{fig5}
\end{figure}

The resulting distribution of time-averaged populations $\langle n_{\text{c}}\rangle$ for a central site is depicted in Fig.~\ref{fig5}, for both one- and two-dimensional lattices and the parameters of Figs.~\ref{figure3}(a) and~\ref{figure4}(b). In addition, we show on top the average value over all trajectories (black solid lines). We observe in both cases that close to the L-R shift of the product density matrix solution, the distributions are centered around two distinct population values for the same detuning $\Delta/\gamma$. Thus the mean-field trajectory simulations also indicate the existence of bistability in the driven-dissipative model.

Note however that the bistable regimes obtained from the product-state Monte Carlo approach are notably more narrow than those shown in Section~\ref{master_eq_mf_section}. In particular, for the 2D case the hysteresis zone has collapsed into a very narrow $\Delta/\gamma$ regime, located at the L-R shift of the product density matrix solution. This is because both methods, although corresponding to a mean-field approximation, are not equivalent. In particular, each individual trajectory $|\Psi^{(r)}(t)\rangle$ leads to a contribution $|\Psi^{(r)}(t)\rangle\langle\Psi^{(r)}(t)|$ to the density matrix of the system, which does not have any classical or quantum correlations. However, by averaging over all $N_{\text{traj}}$ trajectories and $N_{\text{T}}$ contributions at times $t_l$ from each trajectory, we are formally describing the NESS of the system by the density matrix of the form
\begin{equation} \label{combi_trajs}
\rho\propto\sum_{r,\ell} |\Psi^{(r)}(t_\ell) \rangle \langle \Psi^{(r)}(t_\ell) |,
\end{equation}
which does not have a product form as in Eq.~\eqref{mf_ansatz}. Thus in the product density matrix approach all spatial correlations are neglected, while the mean-field trajectory approach fully discards entanglement while retaining other types of correlations~\footnote{Besides classical correlations, the state of Eq.~\eqref{combi_trajs} could still posses a different type of quantum correlations, namely discord, which may exist even in separable mixed states. See Ref.~\cite{vlatko_discord} for a recent review.}. As seen in Fig.~\ref{fig5}, this already has a strong impact on the NESS of the driven-dissipative model. It is then natural to ask whether considering more correlations might eventually suppress the bistable response completely. This question is addressed in Section~\ref{mp_section} by including long-range correlations in the description of the system.

\section{Matrix product description} \label{mp_section}

After analyzing the bistability featured by the mean-field driven-dissipative $XY$ model, we wish to determine whether such a behavior is maintained when spatial quantum correlations are taken into account, or if it is an artifact of the nonlinearity induced by the mean-field description. Furthermore, we expect to observe whether other interesting effects emerge due to these correlations.   

To assess the effect of retaining correlations, we employ an MPO description~\cite{vidal2003efficient,zwolak2004mixed,verstraete04} of the NESS $\rho$ for one dimensional systems. This approach gives an approximate way to account for quantum and classical correlations in the NESS. Intuitively, the parameter $\chi$ controlling the size of the MPO matrices gives a measure of the inter-site correlations, of either classical or quantum origin, so highly correlated states require a larger $\chi$ for an accurate description. In the extreme case $\chi = 1$ the MPO reduces to the mean-field product ansatz of Eq.~\eqref{mf_ansatz}. By solving the NESS with increasing $\chi$ we connect the mean-field approximation with the formally exact but unobtainable limit $\chi \rightarrow \infty$. For each $\chi$ considered, the corresponding MPO density matrix is efficiently evolved in time under Eq.~\eqref{eq:master_equation} using the time evolving block decimation algorithm~\cite{vidal2003efficient,zwolak2004mixed,verstraete04}, where the NESS is obtained by taking the large time limit. We also perform Monte Carlo wavefunction simulations, where each quantum trajectory is calculated within a matrix product state (MPS) description~\cite{vidal2004tebd,schollwock11}. Our implementation of both these methods is based on the open-source Tensor Network Theory (TNT) library~\cite{tnt}.

\subsection{Disappearance of mean-field bistability}

\begin{figure}
\centering
\includegraphics[scale=1.05]{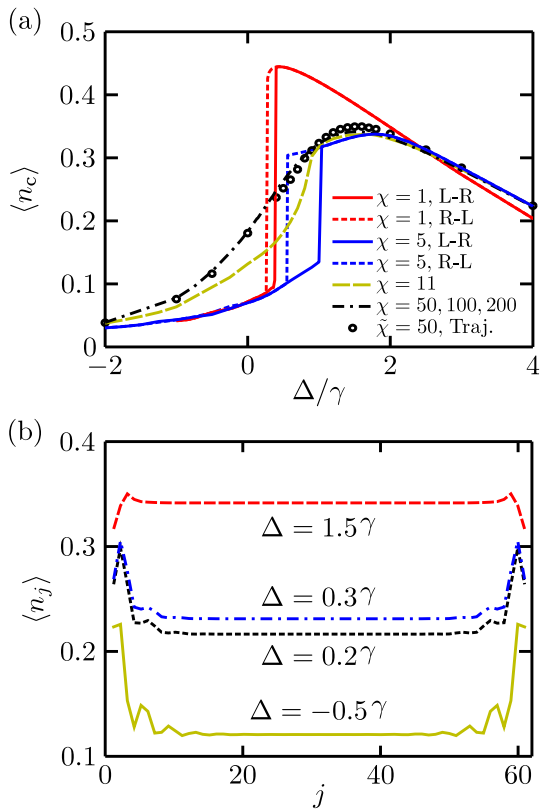}
\caption{(a) Central site density $\langle n_{\text{c}}\rangle$ as a function of $\Delta/\gamma$ for different scenarios, namely L-R and R-L sweeps for MPO approach with $\chi=1$ (i.e. mean-field) and $\chi=5$, a single sweep for $\chi=11,50,100,200$ (where the latter three coincide), and a trajectory simulation for $\tilde{\chi}=50$. For the latter, a time average was performed over the final $80\%$ of the total time of evolution, for which convergence was verified. The results correspond to $J/\gamma=2$, $\Omega/\gamma=1$, and $N=61$. (b) Density profiles for MPOs with $\chi=200$, for the same values of $\Delta/\gamma$ of Fig.~\ref{fig2}(a).}
\label{fig6}
\end{figure}

First we observe what happens to the bistable behavior when increasing the value of $\chi$ in the MPO description of the NESS. As shown in Fig.~\ref{fig6}(a) for the central density $\langle n_{\text{c}}\rangle$, a significant change occurs with respect to the mean-field results when taking $\chi=5$, where just a small amount of correlations is retained across the system. Even though the bistability is still present, it shifts towards larger values of $\Delta/\gamma$, and the high-density regime is notably lower that its mean-field counterpart. Also, the bistability extends over a wider range of $\Delta/\gamma$ values, which might initially suggest that the bistable behavior is strengthened by correlations. However, taking larger values of $\chi$ shows that this is not the case. In fact, for $\chi=11$ we find that the L-R and R-L sweeps give identical NESS, so the bistable behavior has already disappeared. In addition, the shift from the low- to the high-density regime is no longer sharp~\footnote{This result qualitatively differs from those of previous variational studies beyond a mean-field approach of driven-dissipative translationally-invariant 2D lattices with different types of couplings~\cite{Weimer2015_1,Weimer2015_2}. Namely a sharp transition between low- and high-density states is still observed there. Whether this transition smoothens when considering a larger amount of correlations in finite 2D systems remains an open question.}. Further increases of $\chi$ improve the NESS, smoothening the shift between the two density regimes. Finally, from $\chi\approx50$ the NESS remains essentially unchanged with increasing $\chi$. This is indicated in Fig.~\ref{fig6}(a) for the central density, and also for the density profiles in Fig.~\ref{fig6}(b), where the results for $\chi=50,100,200$ coincide. A similarity with the mean-field limit remains though, seen when comparing Fig.~\ref{fig2}(a) with Fig.~\ref{fig6}(b). Namely the high-density regimes also have flat profiles, while the low-density case shows density oscillations that decay when approaching the center of the lattice.

To provide more support to our results, we also obtain the NESS properties of the system from a Monte Carlo wave function approach. In this case we represent the wave function of each independent trajectory by a MPS with maximum matrix size $\tilde{\chi}$ limiting the corresponding amount of quantum entanglement. For each value of $\Delta/\gamma$ considered we simulated at least 10 trajectories. To smoothen the results we also averaged over several hundreds of time steps, resulting in an effective average over thousands of trajectories. As shown in Fig.~\ref{fig6}(a), densities for $\tilde{\chi}=50$ already agree with those of $\chi\geq50$ for a MPO description of the density matrix, and thus further confirm that the bistability is broken when enough correlations are taken into account. 

In summary, we have shown that the physics of the driven-dissipative system obtained from mean-field theory is {\em qualitatively} wrong, with the bistability being an artifact of the nonlinearity induced by the ansatz of Eq.~\eqref{mf_ansatz}. When a MPO or a quantum trajectory MPS description of the NESS are used with large values of $\chi$, and the calculation becomes formally closer to the exact result, the bistability and sharp shift between low- and high-density regimes are washed out by correlations. 

\subsection{Correlations and bunching-antibunching transition}
\begin{figure}
\centering
\includegraphics[scale=1.05]{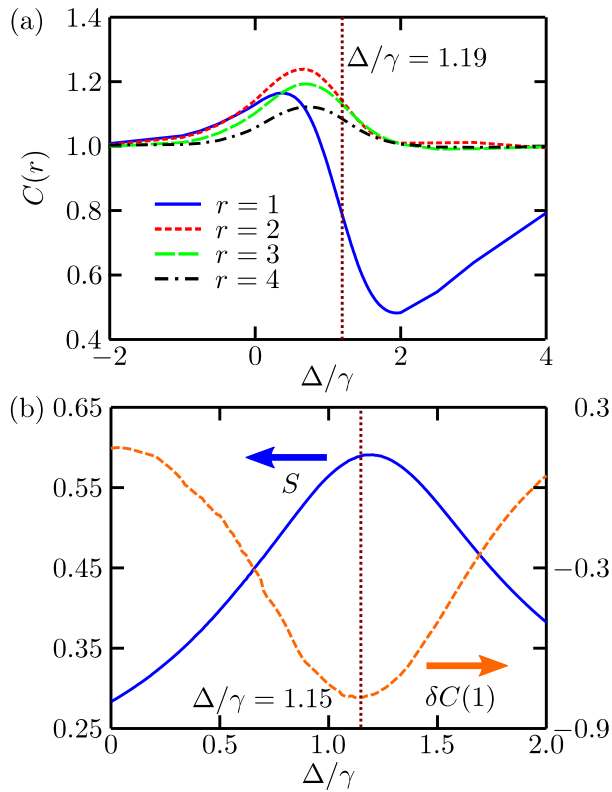}
\caption{(a) Two-site correlations $C(r)$ from the central site of the system and $r=1,2,3,4$ as a function of $\Delta/\gamma$, for $J/\gamma=2$, $\Omega/\gamma=1$, $N=61$ and $\chi=200$. (b) Entropy $S$ and first derivative of the nearest-neighbor normalized correlation $C(1)$ with respect to $\Delta/\gamma$. The locations of the maximum $S$ and minimum $\delta C(1)$ were obtained by fitting the surrounding hill and valley to quadratic functions. In both figures the vertical line indicates the point of fastest decay of correlations $C(1)$.}
\label{fig7}
\end{figure}

Now we show that a new interesting property, not captured by mean-field approaches, emerges in the driven-dissipative system when correlations are taken into account. For this we consider the normalized correlations \footnote{The bunching-antibunching transition can also be observed with the correlations $g^2(j,k)=\frac{\langle\sigma_j^+\sigma_k^+\sigma_j^-\sigma_k^-\rangle}{\langle\sigma_j^+\sigma_j^-\rangle\langle\sigma_k^+\sigma_k^-\rangle}$, which can be rewritten as $g^2(j,k)=\frac{1+\langle\sigma_j^z\rangle+\langle\sigma_k^z\rangle+\langle\sigma_j^z\sigma_k^z\rangle}{(1+\langle\sigma_j^z\rangle)(1+\langle\sigma_k^z\rangle)}$. We take instead the correlations in Eq.~\eqref{norm_correl}, and thus discard the common factor $(1+\langle\sigma_j^z\rangle+\langle\sigma_k^z\rangle)$ in both the numerator and denominator of the $g^2$ correlations.}
\begin{equation} \label{norm_correl}
C(j,r)=\frac{\langle\sigma_j^z\sigma_{j+r}^z\rangle}{\langle\sigma_j^z\rangle\langle\sigma_{j+r}^z\rangle},
\end{equation}
which tend to 1 in the mean-field limit. In Fig.~\ref{fig7}(a) we show the correlations around the center of the system (i.e. for site $j=\lceil N/2\rceil$), simply denoted as $C(r)$, for $r=1,2,3,4$ and $\chi=200$ as a function of $\Delta/\gamma$. We have verified that the same results are obtained for correlations centered around any other site in the bulk of the chain.

First note that, as expected, the normalized correlations tend to the mean-field limit as $\Delta$ becomes the dominant energy scale in the system. Additionally, for $r>1$, they always remain above unity and decrease monotonically as $r$ increases. However, the most important observation corresponds to the correlations $C(1)$, which as depicted in Fig.~\ref{fig7}(a) cross unity at $\Delta/\gamma=0.91$. This indicates a transition from a regime in which neighboring excitations tend to cluster together ($\Delta/\gamma<0.91$) to a configuration where the existence of such clusters is disfavored and excitations tend to spread out ($\Delta/\gamma>0.91$). In other words, our simulations show that the system features a bunching-antibunching transition. As $\Delta/\gamma$ increases $C(1)$ continues decreasing until $\Delta/\gamma=1.94$, where it reaches its minimum value. Then $C(1)$ grows again towards the mean-field limit, as $\Delta/\gamma$ becomes very large.   

Finally we calculate an alternative quantity which measures the amount of both classical and quantum correlations, namely the entropy~\cite{prosen_zni2009,mendoza2013}
\begin{equation}
S=-\sum_{\alpha}\lambda_{\alpha}^2\log_2\lambda_{\alpha}^2,
\end{equation}
calculated from the Schmidt coefficients $\lambda_{\alpha}$ which result when the full NESS density matrix $\rho$ is factorized into two half chains as
\begin{equation}
\rho=\sum_{\alpha}\lambda_{\alpha}O_{\alpha}^AO_{\alpha}^B,
\end{equation}
and normalized so the first coefficient is $\lambda_1=1$. Note that for the mean-field product density matrix $S=0$. In Fig.~\ref{fig7}(b) we show the entropy, together with the first derivative of the nearest-neighbor normalized correlation $C(1)$ with respect to $\Delta/\gamma$, denoted as $\delta C(1)$. The maximal entropy occurs at $\Delta/\gamma=1.19$, which is very close to the point of fastest decay of $C(1)$, namely at $\Delta/\gamma=1.15$. This correspondence between maximal entropy and fastest immersion within the antibunched phase indicates that the transition is indeed a strong-correlation effect.

\section{Physical implementation in circuit QED} \label{simulation_section}
We have shown that in 1D the bistability observed in a mean-field description of the driven-dissipative $XY$ model is broken when enough correlations are taken into account. This result agrees with previous calculations based on TNT methods and on a recently-proposed technique which introduces correlations among neighboring sites of the lattice~\cite{Degenfeld2014}. In addition, we have shown that a bunching-antibunching transition emerges instead. However the physics in higher dimensions is not fully understood. First, unlike equilibrium systems, there is no reason to believe that mean-field theory will be more accurate despite the higher coordination number. Second, in spite of important recent developments in TNT methods for studying two-dimensional quantum lattices~\cite{verstraete08,cirac09,orus13,stoudenmire12,yan11,lubasch14}, classical simulations of sufficiently large driven open quantum systems to meaningfully compare results to mean field theory are currently out of reach. Our mean-field Monte Carlo calculations in 2D driven-dissipative lattices already indicate a large impact of the retained correlations on the bistable behavior. Moreover the methods of Refs.~\cite{Weimer2015_1,Degenfeld2014} indicate that beyond mean-field, correlations among neighbors break the bistability in 2D; however determining the impact of long-range correlations is beyond their capabilities. Thus, gaining key insight into driven-dissipative quantum systems by determining their behavior in higher dimensions is a compelling reason to instead develop a quantum simulator~\cite{johnson14,Georgescu2014,dalmonte2015}. In addition, this would provide an experimental setup to confirm the one-dimensional effects discussed in Section~\ref{mp_section}, and to explore other interesting effects found in correlated driven-dissipative models such as dynamic hysteresis~\cite{casteels2015}. In the following we propose a platform for this quantum simulator, accessible using current technology.


\subsection{Circuit QED with transmon qubits}
\begin{figure}[t]
\centering
\includegraphics[scale=1.1]{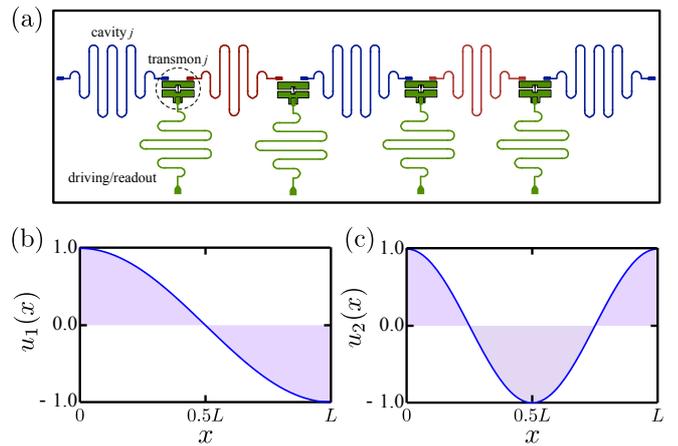}
\caption{\label{XYcQED} (a) Circuit QED configuration that involves an array of transmon qubits and coplanar waveguide resonators (cavities). Each cavity (blue/red) is coupled to two transmon qubits via electrostatic interaction. In addition, each transmon is coupled to an additional cavity (green) for manipulating and readout of the transmon state. (b) Spatial distribution of the first electromagnetic mode supported by an extended superconducting cavity. (c) Spatial distribution of the second electromagnetic mode.}
\end{figure}

Circuit quantum electrodynamics (QED)~\cite{Blais2004,Wallraff2004,Chiorescu2004}, which involves the interaction between on-chip coplanar waveguides resonators (CWR) and superconducting qubits made of Josephson junctions, represents a prime candidate to simulating many-body physics~\cite{angelakis2007photon,hartmann2006strongly,greentree2006quantum,DWave,Houck2012,Koch2013}. Furthermore, recent experimental achievements may pave the way to the implementation of complex and scalable arrays of superconducting circuits~\cite{Steffen2013}. In this sense, one could implement driven-dissipative many-body dynamics by means of an array of several transmon qubits~\cite{Koch2007} coupled to CWR, as depicted in Fig.~\ref{XYcQED}(a). Here, each coplanar waveguide resonator or cavity (blue/red) interacts via electrostatic energy with two transmon qubits, and there is no direct transmon-transmon interaction. In addition, each transmon is coupled to an additional cavity (green) which is used to manipulate the qubit state via classical microwaves, as well as qubit readout. In what follows, we present the basic tools to simulating the ferromagnetic/antiferromagnetic $XY$ model.                   

First, each CWR represents an extended superconducting device which supports a discrete number of electromagnetic modes determined by specific boundary conditions. In our case, we need open boundary conditions such that the voltage distribution at the cavity edges is a maximum. The quantization of an extended cavity can be found elsewhere~\cite{Blais2012}, so we will present the main results. The CWR can be described by the voltage distribution
\begin{equation}
V(x,t) = i\sum_n \Bigg(\frac{\hbar\omega_n}{2C_r}\Bigg)^{1/2}(a^{\dag}_n-a_n)u_n(x),
\end{equation}
where $a_n$ ($a^{\dag}_n$) is the annihilation (creation) bosonic operator, $\omega_n$ is the $n$th cavity frequency, and $C_r$ is the total capacitance of the cavity. The eigenfunction $u_n(x)=A_n\cos(k_nx)$ takes into account the spatial distribution of the cavity with wave vectors defined by $k_n=n\pi/L$ ($n \in {Z}^+$), and $L$ is the cavity length. The spatial distribution for the first ($n=1$) and second ($n=2$) cavity modes are shown in Figs.~\ref{XYcQED}(b) and~\ref{XYcQED}(c) respectively.

Second, the electrostatic interaction between a transmon and a coplanar waveguide resonator reads~\cite{Koch2007}
\begin{equation}
H_{\rm int} = 2e\beta\hat{n}V(x,t),
\end{equation}
where $e$ is the electron charge, $\beta$ is a dimensionless parameter, and $\hat{n}$ is the Cooper-pair number in the superconducting island which defines the transmon device. Because of the slight anharmonicity of the transmon spectrum~\cite{Koch2007}, we can control and define a TLS or qubit interacting with a single mode of the electromagnetic field via the Jaynes-Cummings interaction
\begin{equation}
H = \omega_0\sigma^{+}\sigma^- + \omega a^{\dag}a + g(\sigma^{+}a + \sigma^{-}a^{\dag}).
\end{equation}
Notice that the sign of the qubit-cavity coupling strength $g$ depends on the position of the transmon along the cavity, and the specific mode that we choose to work with, see Figs.~\ref{XYcQED}(b) and~\ref{XYcQED}(c). This is our starting point for the simulation of the ferromagnetic/antiferromagnetic $XY$ model in circuit QED.

\subsection{Quantum simulation of the $XY$ model}
Let us consider the situation depicted in Fig.~\ref{XYcQED}(a), where we assume identical transmon qubits with energy $\omega_c$, and cavities with frequencies such that $\omega_j=\omega_{j+2}$ and $\omega_{j}\neq\omega_{j+1}$. The above condition can be satisfied for cavities with different lengths, as represented in our scheme with blue and red cavities. In addition, we consider that each qubit is manipulated by a classical microwave of amplitude $\Omega$ and driving frequency $\omega_L$. In this case, the Hamiltonian that describes the quantum dynamics is
\begin{align}
\begin{split}
H&=\omega_{c}\sum_j \sigma^{+}_j\sigma_j^- + \sum_{j}\omega_{j} a^{\dag}_{j}a_{j}\\
&+ \Omega\sum_j(\sigma^{+}_je^{-i\omega_Lt}+\sigma^{-}_je^{i\omega_Lt}) \\
&+\sum_{j}[g_{j+1}(\sigma^{+}_ja_{j+1}+\sigma^{-}_ja^{\dag}_{j+1})\mp g_{j}(\sigma^{+}_ja_{j}+\sigma^{-}_ja^{\dag}_{j})],
\label{GeneralHamiltonian}
\end{split}
\end{align}
where the coupling strengths satisfy the conditions $g_j=g_{j+2}$ and $g_j\neq g_{j+1}$. The qubits are described by Pauli transition matrices $\sigma^{\pm}_j$, and the bosonic fields by creation and annihilation operators $a^{\dag}_j, a_j$. The $-$ and $+$ signs that appear in the qubit-cavity interaction come from the choice of the first and second mode of each cavity, respectively. As we show below, they result in the simulation of ferromagnetic and antiferromagnetic $XY$ models respectively.

The $XY$ model can be implemented if we consider the dispersive regime such that virtual photons provide the direct qubit-qubit coupling. In a reference frame rotating with the driving frequency $\omega_L$, the interaction Hamiltonian reads
\begin{align}
&H_I(t)=\Delta_c\sum_j\sigma^+_j\sigma^-_j-\sum_j\Delta_ja^{\dag}_ja_j+\Omega\sum_j(\sigma_j^{+}+\sigma_j^{-})\nonumber\\
&+\sum_{j}[g_{j+1}(\sigma^{+}_ja_{j+1}+\sigma^{-}_ja^{\dag}_{j+1})\mp g_{j}(\sigma^{+}_ja_{j}+\sigma^{-}_ja^{\dag}_{j})],
\label{Hint}
\end{align}
where we define the detunings $\Delta_c=\omega_c-\omega_L$ and $\Delta_j=\omega_L-\omega_j$, the latter satisfying the conditions $\Delta_j=\Delta_{j+2}$,  $\Delta_j\neq\Delta_{j+1}$ according to the previous definition of frequencies. One can access the dispersive regime if the condition $|\Delta_{j}|\gg \{g_{j},\Delta_c\}$ is satisfied. For example, consider an array of three transmon qubits and four cavities. In this case, the effective second-order Hamiltonian is
\begin{align}
&H_{\rm eff}=\Big(\Delta_c + \frac{g^2_1}{\Delta_1}+\frac{g^2_2}{\Delta_2}\Big)\sum_j\sigma^{+}_j\sigma_j^-+\Omega\sum_j(\sigma_j^{+}+\sigma_j^-) \nonumber\\
&+\frac{1}{\Delta_1}\left(2g^2_1\sigma^{+}_1\sigma_1^--(g_1^2+\Delta_1^2)\right)a^{\dag}_1a_1\nonumber\\
&+\frac{1}{\Delta_2}\left(2g^2_2(\sigma^{+}_1\sigma_1^-+\sigma^{+}_2\sigma_2^-)-(2g_2^2+\Delta_2^2)\right)a^{\dag}_2a_2 \nonumber\\
&+\frac{1}{\Delta_1}\left(2g^2_1(\sigma^{+}_2\sigma_2^-+\sigma^{+}_3\sigma_3^-)-(2g_1^2+\Delta_1^2)\right)a^{\dag}_3a_3 \nonumber\\
&+\frac{1}{\Delta_2}\left(2g^2_2\sigma^{+}_3\sigma_3^--(g_2^2+\Delta_2^2)\right)a^{\dag}_4a_4 \nonumber\\
&\mp\frac{g^2_1}{\Delta_1}(\sigma^{+}_1\sigma^{-}_2 + \sigma^{-}_1\sigma^{+}_2) \mp\frac{g^2_2}{\Delta_2}(\sigma^{+}_2\sigma^{-}_3 + \sigma^{-}_2\sigma^{+}_3)\nonumber\\
&+\frac{g_1g_2}{2}\Big(\frac{1}{\Delta_1} + \frac{1}{\Delta_2}\Big)(a^{\dag}_1a_2e^{i(\Delta_1-\Delta_2)t} + {\rm H.c.})\nonumber\\
&+\frac{g_1g_2}{2}\Big(\frac{1}{\Delta_1} + \frac{1}{\Delta_2}\Big)(a^{\dag}_2a_3e^{i(\Delta_2-\Delta_1)t} + {\rm H.c.}).
\label{Heff}
\end{align}

This Hamiltonian can implement the ferromagnetic/antiferromagnetic $XY$ model if there are no photons initially present in the dynamics, and if the condition $|\Delta_1-\Delta_2|\gg J_{12},J_{23}$ is satisfied, where 
\begin{align}
J_{12}=\frac{g_1g_2}{2} (\frac{1}{\Delta_1} + \frac{1}{\Delta_2}),\quad J_{23}=\frac{g_2g_3}{2} (\frac{1}{\Delta_1} + \frac{1}{\Delta_2}), 
\end{align}
and $g^2_2=g^2_1(\Delta_2/\Delta_1)$. In addition, the Stark shifts associated to each qubit can be suppressed by changing the qubit frequencies, which can be achieved by the application of an external flux on each transmon~\cite{Koch2007}. The extension to a large number of transmon qubits and cavities is straightforward. 

\begin{figure}[t]
\centering
\includegraphics[scale=1.05]{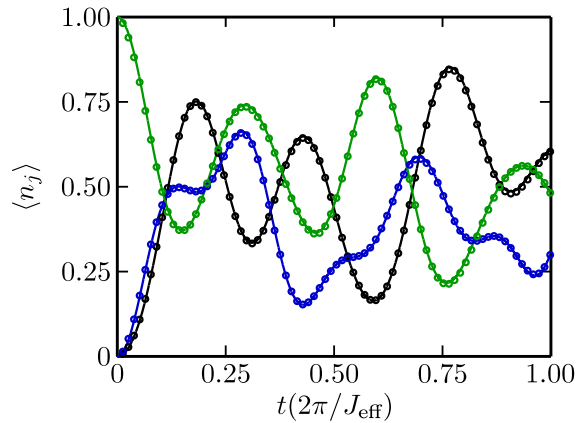}
\caption{\label{XYcQED1} Expectation value of operator $n_j=(1+\sigma_j^z)/2$ for $j=1$ (green), $j=2$ (blue), and $j=3$ (black). The continuous lines correspond to the dynamics governed by the exact ferromagnetic $XY$ model, while the circles ($\circ$) correspond to the simulated model. 
Here we have fixed the detuning parameters in units of the qubit-cavity coupling $g_1$, namely $\Delta_1=30g_1$ and $\Delta_2=20g_1$, and have taken $g_1=1$, which results in an effective coherent tunneling rate $J_{\rm eff}=g_1^2/\Delta_1=0.0333$. We have also chosen $\Delta_c=\Omega=J_{\rm eff}$.} 
\end{figure}

We have performed numerical simulations starting from the Hamiltonian of Eq.~\eqref{Hint} (ferromagnetic case) to test our approach, and then compared it to the exact $XY$ model. In Fig.~\ref{XYcQED1} we depict the expectation value of the population $n_j$ for each qubit $j$, without including decay processes. The results show a quite good matching between the simulated ferromagnetic $XY$ model ($\circ$) and the exact model (solid lines) if we change
\begin{equation}
\Delta_c\rightarrow\Delta-\frac{g_1^2}{\Delta_1}-\frac{g_2^2}{\Delta_2},
\end{equation}
where $\Delta$ is the simulated detuning that appears in Eq.~\eqref{hami_first}.

\subsection{Realistic parameter regime}
The main features of the driven-dissipative many-body system appear in a well defined range of system parameters. In terms of the decay rate of qubits $\gamma$, the driving amplitude $\Omega$ ranges from $0.1\gamma$ to $2\gamma$, the detuning $\Delta_c$ from $-2\gamma$ to $10\gamma$, and the coherent tunneling rate $J$ from $0$ to $10\gamma$. It is important to mention that in a realistic scenario the qubits experience relaxation and dephasing with typical coherence times of about $T_1\sim 1\mu$s and $T_2\sim 0.6 \mu$s~\cite{Leo2010}. In the latter experiment, the qubits have frequencies $\omega^{(1)}_c/2\pi=6$~GHz $\pm 2$~MHz, $\omega^{(2)}_c/2\pi=7$~GHz $\pm 2$~MHz, and $\omega^{(3)}_c/2\pi=8$~GHz $\pm 2$~MHz. The coherence time $T_1$ gives a relaxation rate $\gamma_1\sim 1$~MHz, so $\Omega$ ranges from $0.1$~MHz to $2$~MHz, $\Delta_c$ from $-2$~MHz to $10$~MHz, and $J$ from $0$ to $10$~MHz.

The above parameter regimes can be obtained with state-of-the-art circuit QED technology. The driving amplitude $\Omega$ is limited by the cryostat cooling power and it may range from $0$ to $2\pi\times 0.7$~GHz~\cite{Ballester2012}. The driving frequency $\omega_L$ may range from $0$ to $2\pi\times 18$~GHz. The effective coherent tunneling rate $J_{\rm eff}$, can be tuned from zero to a maximal value if we consider a transmon with Purcell protection and tunable qubit-cavity coupling~\cite{Blais2011}. In addition, if the maximal qubit-cavity coupling is about $g_1/2\pi\sim100$~MHz, the detuning $\Delta_1$ in Eq.~(\ref{Hint}) must be $\Delta_1\sim 2\pi \times 6$~GHz in order to reach $J_{\rm eff}=10$~MHz. This value of $\Delta_1$ is attainable since the resonator frequency $\omega_1/2\pi$ ranges from $2$~GHz to $10$~GHz.

\section{Conclusions} \label{conclu}
In the present work we have discussed how a mean-field product ansatz, while being qualitatively successful in describing equilibrium settings, can lead to dubious physical conclusions for driven-dissipative nonequilibrium systems. In particular, we have shown that single-site mean-field calculations of the NESS of a coherently-driven $XY$ model with local dissipation predict bistable behavior. This is manifested in the existence of two different history-dependent values of a critical driving field detuning $\Delta/\gamma$ at which a sharp shift between low- and high-density regimes occurs. However, when the dynamics of a one-dimensional lattice is simulated with a matrix-product description, which considers both classical and quantum correlations, the bistability disappears along with the sharp shift between the two different density states. Instead, the density profile becomes smooth, and a bunching-antibunching transition, which cannot be determined from a mean-field approach, emerges in the nearest-neighbor correlations.  

Our findings highlight the challenges in describing non-equilibrium quantum lattice systems and motivate their quantum simulation, most specially in higher spatial dimensions where accurate classical simulations are not currently feasible. With this in mind, we have elucidated a possible experimental realization of such a quantum simulator using transmon qubits in a circuit QED, which can be implemented using current technology.

\begin{acknowledgments}
We acknowledge funding by the ERC under the European Union's Seventh Framework Programme (FP7/2007-2013)/ERC Grant Agreement no 319286, Q-MAC; UK EPSRC funding EP/K038311/1; the National Research Foundation and the Ministry of Education of Singapore; the Spanish grant MINECO FIS2012-36673-C03-02; UPV/EHU UFI 11/55; Basque Government Grant IT472-10; PROMISCE and SCALEQIT EU projects; and The Fondo Nacional de Desarrollo Cient\'ifico y Tecnol\'ogico (FONDECYT, Chile) under Grant 1150653. D.J. acknowledges financial support from the EU Collaborative project QuProCS (Grant Agreement 641277). J.J.M.-A. acknowledges T. Grujic for his collaboration in the early stages of the project, S. Caballero for his comments on the manuscript, and L. Quiroga and F. Rodr\'{i}guez for their hospitality at Universidad de los Andes.
\end{acknowledgments}

\appendix
		
\section{Mean-field approximation for master equation dynamics} \label{mfa_me}

Here we present the mean field approximation used to study the dynamics and NESS of the system. 
We describe the total time-dependent state of the system $\rho(t)$ of $N$ sites by means of the ansatz of Eq.~\eqref{mf_ansatz},
where the reduced density operator of site $j$ is given by 
\begin{equation} \label{matrix_rho}
\rho_j(t)=\begin{pmatrix}
\rho^{11}_j(t) & \rho^{10}_j(t) \\[6pt]
\rho^{01}_j(t) & \rho^{00}_j(t)
\end{pmatrix}.
\end{equation}
Here $\rho^{11}_j(t)$ corresponds to the population of the ``up" state of site $j$ at time $t$, $\rho^{00}_j(t)$ to the population of the ``down" state, and $\rho^{10}_j(t)$ and $\rho^{10}_j(t)$ are the coherences. We insert Eq.~\eqref{mf_ansatz} into the master equation describing the dynamics of the system, namely
\begin{equation}
\frac{d\rho}{dt}=-i[\mathcal{H},\rho]+\mathcal{D}(\rho),
\end{equation}
where $\mathcal{H}$ is the total Hamiltonian and $\mathcal{D}(\rho)$ is the dissipative component. Tracing out all degrees of freedom except those of site $j$ (which we denote by $\text{Tr}(\ldots)_{j'}$), we obtain the equation for the reduced density operator of site $j$,
\begin{equation}
\frac{d\rho_j}{dt}=\text{Tr}\left(\frac{d\rho_j}{dt}\right)_{j'}=-i[\mathcal{H}_j^{\text{mf}},\rho_j]+\mathcal{D}(\rho_j),
\end{equation}
with the local mean field Hamiltonian
\begin{align} \label{mean_field_hamiltonian}
\begin{split}
\mathcal{H}_j^{\text{mf}}=\Delta\sigma_j^+\sigma_j^-&+\biggl[\Omega-J\sum_{k}\langle\sigma_{k}^-\rangle\biggr]\sigma_j^++\biggl[\Omega-J\sum_{k}\langle\sigma_{k}^+\rangle\biggr]\sigma_j^-\\=\Delta\sigma_j^+\sigma_j^-&+\Omega_j\sigma_j^++\Omega_j^*\sigma_j^-.
\end{split}
\end{align}
with the sums performed over the nearest neighbors $k$ of site $j$, effective driving amplitudes $\Omega_j$ as defined in Eq.~\eqref{effective_driving_mf}, and local dissipator
\begin{equation} \label{incoherent_obc}
\mathcal{D}(\rho_j)=\gamma\left(\sigma_j^-\rho_j\sigma_j^+-\frac{1}{2}\sigma_j^+\sigma_j^-\rho_j-\frac{1}{2}\rho_j\sigma_j^+\sigma_j^-\right).
\end{equation}
This leads to a nonlinear set of equations for the components $\rho^{\alpha\beta}_j(t)$ of each local reduced density operator, which depend on expectation values of neighboring sites. Defining $\vec{\rho}_j=(\rho^{00}_j,\rho^{01}_j,\rho^{11}_j,\rho^{10}_j)^{\text{T}}$, this set of equations is given by
\begin{equation} \label{local_mf_eq}
\frac{d}{dt}\vec{\rho}_j=\mathcal{L}_{j}^{\text{mf}}\vec{\rho}_j,\qquad\text{with}
\end{equation} 
\begin{equation} \label{mf_local_evolution_obc}
\mathcal{L}_j^{\text{mf}}=\begin{pmatrix}
0 & i\Omega_j & \gamma & -i\Omega_j^*\\
i\Omega_j^* & i\Delta-\frac{\gamma}{2} & -i\Omega_j^* & 0\\
0 & -i\Omega_j & -\gamma & i\Omega_j^*\\
-i\Omega_j & 0 & i\Omega_j & -i\Delta-\frac{\gamma}{2}
\end{pmatrix}.
\end{equation} 
The total evolution of the system is thus calculated by evaluating the evolution of each site during small time intervals of length $\delta t$, using expectation values of the immediately previous time, namely
\begin{equation}
\vec{\rho}_j(t+\delta t)=e^{\mathcal{L}_j^{\text{mf}}(t)\delta t}\vec{\rho}_j(t).
\end{equation}
Evolving for a very long time, until the disappearance of the transient dynamics, we obtained the mean-field NESS of the system discussed in Section~\ref{master_eq_mf_section}.

\section{Mean-field approximation for Monte Carlo wave function method} \label{mfa_tm}
As described in the main text, an alternative way to analyze the physics of driven-dissipative systems corresponds to the simulation of several independent stochastic trajectories, whose average in the long time limit gives the NESS. 

Here we briefly mention a few points of the method in the mean-field limit, used in Section~\ref{trajectories_section} to show the existence of bistability in the absence of quantum correlations. To stay in this limit we assume that at every time step the pure state of each trajectory is given by a product of the form in Eq.~\eqref{mf_pure_state}. It then follows that between the application of jump operators, the evolution of the full lattice for trajectory $r$ can be performed by evolving each site separately at each time step, as
\begin{equation}
|\psi^{(r)}_j(t+\delta t)\rangle=e^{-i\mathcal{H}_j^{\text{eff}(r)}(t)\delta t}|\psi^{(r)}_j(t)\rangle,
\end{equation}
with effective mean-field Hamiltonian 
\begin{equation}
\mathcal{H}_j^{\text{eff}(r)}(t)=\mathcal{H}_j^{\text{mf}(r)}(t)-\frac{i}{2}\gamma\sigma_j^{+}\sigma_j^-,
\end{equation}
where the imaginary term results from the coupling of the lattice to the environment. We have explicitly pointed out the time and trajectory dependence of the effective Hamiltonian, coming from the dependence of $\mathcal{H}_j^{\text{mf}}$ on neighboring expectation values, which are different for each trajectory due to the random application of jump operators across the time evolution.


\begin{thebibliography}{10}

\bibitem{kurth2010}
S. Kurth, G. Stefanucci, E. Khosravi, C. Verdozzi and E. K. U. Gross, Phys. Rev. Lett. {\bf 104}, 236801 (2010).

\bibitem{wilner2013}
E.Y. Wilner, H. Wang, G. Cohen, M. Thoss and E. Rabani, Phys. Rev. B {\bf 88}, 045137 (2013).

\bibitem{khosravi2012}
E. Khosravi, A.-M. Uimonen, A. Stan, G. Stefanucci, S. Kurth, R. van Leeuwen and E.K.U. Gross, Phys. Rev. B {\bf 85}, 075103 (2012).

\bibitem{benenti2009negative}
G.~Benenti, G.~Casati, T.~Prosen, and D.~Rossini, Europhys. Lett. {\bf 85}, 37001 (2009).

\bibitem{vznidarivc2011spin}
M.~{\v{Z}}nidari{\v{c}}, Phys. Rev. Lett. {\bf 106}, 220601 (2011).

\bibitem{prosen_zni2009}
T. Prosen and M.~{\v{Z}}nidari{\v{c}}, J. Stat. Mech. P02035 (2009).

\bibitem{mendoza2013}
J. J. Mendoza-Arenas, T. Grujic, D. Jaksch and S. R. Clark, Phys. Rev. B {\bf 87}, 235130 (2013).

\bibitem{mendoza2013jstat}
J. J. Mendoza-Arenas, S. Al-Assam, S. R. Clark, and D. Jaksch, J. Stat. Mech. P07007 (2013).

\bibitem{ssn2010signatures}
A.~Tomadin, V.~Giovannetti, R.~Fazio, D.~Gerace, I.~Carusotto, H.E. T{\"u}reci, and A.~Imamo{\u{g}}lu, Phys. Rev. A {\bf 81}, 061801 (2010).

\bibitem{liu2011quantum}
K.~Liu, L.~Tan, C.H. Lv, and W.M. Liu, Phys. Rev. A {\bf 83}, 063840 (2011).

\bibitem{tomadin2010prl}
S.~Diehl, A.~Tomadin, A. Micheli, R. Fazio, and P.~Zoller, Phys. Rev. Lett. {\bf 105}, 015702 (2010).

\bibitem{tomadin2011nonequilibrium}
A.~Tomadin, S.~Diehl, and P.~Zoller, Phys. Rev. A {\bf 83}, 013611 (2011).

\bibitem{grujic2012}
T. Grujic, S.R. Clark, D. Jaksch and D.G. Angelakis, New J. Phys. {\bf 14}, 103025 (2012).

\bibitem{joshi2013}
C. Joshi, F. Nissen and J. Keeling, Phys. Rev. A {\bf 88}, 063835 (2013).
 
\bibitem{grujic2013}
T. Grujic, S.R. Clark, D. Jaksch and D.G. Angelakis, Phys. Rev. A {\bf 87}, 053846 (2013).

\bibitem{jin2013photon}
J.~Jin, D.~Rossini, R.~Fazio, M.~Leib, and M.J. Hartmann, Phys. Rev. Lett. {\bf 110}, 163605 (2013).

\bibitem{LeBellac04}
M. Le Bellac, F. Mortessagne and G.G. Batrouni, {\em Equilibrium and Non-Equilibrium Statistical Thermodynamics} (Cambridge University Press, Cambridge, 2004). 

\bibitem{Sachdev11}
S. Sachdev, {\em Quantum Phase Transitions, Second Edition} (Cambridge University Press, Cambridge, 2011).

\bibitem{Dombi13}
A. Dombi, A. Vukics and P. Domokos, J. Phys. B: At. Mol. Opt. Phys. {\bf 46}, 224010 (2013).

\bibitem{lee2011}
T. E. Lee, H. H\"affner and M. C. Cross, Phys. Rev. A {\bf 84}, 031402(R) (2011).

\bibitem{boite2012steady}
A.~Le~Boit{\'e}, G.~Orso, and C.~Ciuti, Phys. Rev. Lett. {\bf 110}, 233601 (2013).

\bibitem{nissen2012nonequilibrium}
F.~Nissen, S.~Schmidt, M.~Biondi, G.~Blatter, H.E. T{\"u}reci, and J.~Keeling, Phys. Rev. Lett. {\bf 108}, 233603 (2012).

\bibitem{carr2013}
C. Carr, R. Ritter, C. G. Wade, C. S. Adams, and K. J. Weatherill, Phys. Rev. Lett. {\bf 111}, 113901 (2013).

\bibitem{Fazio2014}
J. Jin, D. Rossini, M. Leib, M.J. Hartmann, R. Fazio, Phys. Rev. A {\bf 90}, 023827 (2014).

\bibitem{LeBoite2014_PRA}
A. Le Boit\'e, G. Orso and C. Ciuti, Phys. Rev. A {\bf 90}, 063821 (2014).

\bibitem{marcuzzi2014}
M. Marcuzzi, E. Levi, S. Diehl, J. P. Garrahan and I. Lesanovsky, Phys. Rev. Lett. {\bf 113}, 210401 (2014).

\bibitem{Naether2015}
U. Naether, F. Quijandr\'{i}a, J. J. Garc\'{i}a-Ripoll and D. Zueco, Phys. Rev. A {\bf 91}, 033823 (2015).

\bibitem{everest2015}
B. Everest, M. Marcuzzi and I. Lesanovsky, arXiv:1507.06909 (2015).

\bibitem{Degenfeld2014}
P. Degenfeld-Schonburg and M. J. Hartmann, Phys. Rev. B {\bf 89}, 245108 (2014).

\bibitem{Weimer2015_1}
H. Weimer, Phys. Rev. Lett. {\bf 114}, 040402 (2015).

\bibitem{Weimer2015_2}
H. Weimer, Phys. Rev. A {\bf 91}, 063401 (2015).

\bibitem{Weimer2015_3}
V. R. Overbeck and H. Weimer, Phys. Rev. A {\bf 93}, 012106 (2016).

\bibitem{finazzi2015}
S. Finazzi, A. Le Boit\'e, F. Storme, A. Baksic, and C. Ciuti, Phys. Rev. Lett. {\bf 115}, 080604 (2015).

\bibitem{Cui2015}
J. Cui, J. I. Cirac and M. C. Ba\~nuls, Phys. Rev. Lett. {\bf 114}, 220601 (2015).

\bibitem{Mascarenhas2015}
E. Mascarenhas, H. Flayac, and V. Savona, Phys. Rev. A {\bf 92}, 022116 (2015).

\bibitem{gardiner04}
C. W. Gardiner and P. Zoller, {\em Quantum Noise}, (Springer-Verlag, 2004).

\bibitem{breuer2002}
H.-P. Breuer and F. Petruccione, {\em The Theory of Open Quantum Systems} (Oxford University Press, Oxford, 2002).

\bibitem{Spohn1977}
H. Spohn, Lett. Math. Phys. {\bf 2}, 33 (1977).

\bibitem{maghrebi2015}
M. F. Maghrebi, and A. V. Gorshkov, arXiv:1507.01939 (2015).

\bibitem{Bonnes2014}
L. Bonnes, D. Charrier, and A. M. L\"auchli, Phys. Rev. A {\bf 90}, 033612 (2014).

\bibitem{glendinning1994stability}
P.~Glendinning, {\em Stability, instability and chaos: an introduction to the theory of nonlinear differential equations} (Cambridge Univ. Press, 1994).

\bibitem{zwolak2004mixed}
M.~Zwolak and G.~Vidal, Phys. Rev. Lett. {\bf 93}, 207205 (2004).

\bibitem{verstraete04}
F. Verstraete, J. J. Garcia-Ripoll and J. I. Cirac, Phys. Rev. Lett. {\bf 93}, 207204 (2004).

\bibitem{vidal2003efficient}
G.~Vidal, Phys. Rev. Lett. {\bf 91}, 147902 (2003).

\bibitem{vidal2004tebd}
G.~Vidal, Phys. Rev. Lett. {\bf 93}, 040502 (2004).

\bibitem{schollwock11}
U. Schollw{\"o}ck, Ann. Phys. {\bf 326}, 96 (2011).

\bibitem{tnt}
S. Al-Assam, S. R. Clark, D. Jaksch, and TNT Development Team. TNT Library Alpha Version, http://www.tensornetworktheory.org, (2014).

\bibitem{verstraete08}
F. Verstraete, J.I. Cirac and V. Murg, Adv. Phys. {\bf 57}, 143 (2008).

\bibitem{cirac09}
J. I. Cirac and F. Verstraete, J. Phys. A: Math. Theor. {\bf 42}, 504004 (2009).

\bibitem{orus13}
R. Or{\'u}s, Ann. Phys. {\bf 349}, 117 (2014).

\bibitem{evenbly11}
G. Evenbly and G. Vidal, Chapter 4 in {\em Strongly Correlated Systems. Numerical Methods}, edited by A. Avella and F. Mancini (Springer Series in Solid-State Sciences, Vol. 176, 2013).

\bibitem{biamonte11}
J. D.  Biamonte, S. R. Clark, D. Jaksch AIP Advances {\bf 1}, 042172 (2011).

\bibitem{plenio98}
M. B. Plenio and P. L. Knight, Rev. Mod. Phys. {\bf 70}, 101 (1998).

\bibitem{Daley2014traj}
A. J. Daley, Adv. Phys. {\bf 63}, 77 (2014).

\bibitem{Buca2012}
B. Bu{\v{c}}a and T. Prosen, New J. Phys. {\bf 14}, 073007 (2012).

\bibitem{Bonnes2014_arxiv}
L. Bonnes and A. M. L\"auchli, arXiv:1411.4831 (2014).

\bibitem{vlatko_discord}
K. Modi, A. Brodutch, H. Cable, T. Paterek, and V. Vedral. Rev. Mod. Phys. {\bf 84}, 1655 (2012).

\bibitem{stoudenmire12}
E.M. Stoudenmire, S. R. White, Annu. Rev. Condens. Matter Phys.  {\bf 3}, 111 (2012).

\bibitem{yan11}
S. Yan, D. A. Huse, S. R. White, Science {\bf 332}, 1173 (2011).

\bibitem{lubasch14}
M. Lubasch, J. I. Cirac, and M.-C. Ba\~{n}uls, Phys. Rev. B {\bf 90}, 064425 (2014).

\bibitem{johnson14}
T. H. Johnson, S. R. Clark and D. Jaksch, Eur. Phys. J. Quant. Technol. {\bf 1}, 1 (2014).

\bibitem{Georgescu2014}
I. M. Georgescu, S. Ashhab, and Franco Nori, Rev. Mod. Phys. {\bf 86}, 153 (2014).

\bibitem{dalmonte2015}
M. Dalmonte, S. I. Mirzaei, P. R. Muppalla, D. Marcos, P. Zoller, and G. Kirchmair, Phys. Rev. B {\bf 92}, 174507 (2015).

\bibitem{casteels2015}
W. Casteels, F. Storme, A. Le Boit{\'e} and C. Ciuti, arXiv:1509.02118 (2015).

\bibitem{Blais2004} 
A. Blais, R. S. Huang, A. Wallraff, S. M. Girvin, and R. J. Schoelkopf, Phys. Rev. A {\bf 69}, 062320 (2004).

\bibitem{Wallraff2004} 
A. Wallraff, D. I. Schuster, A. Blais, L. Frunzio, R.-S. Huang, J. Majer, S. Kumar, S. M. Girvin, and R. J. Schoelkopf, Nature (London) {\bf 431}, 162 (2004).

\bibitem{Chiorescu2004} 
I. Chiorescu, P. Bertet, K. Semba, Y. Nakamura, C. J. P. M. Harmans, and J. E. Mooij, Nature (London) {\bf 431}, 159 (2004).

\bibitem{angelakis2007photon}
D.G. Angelakis, M.F. Santos, and S.~Bose, Phys. Rev. A {\bf 76}, 031805 (2007).

\bibitem{hartmann2006strongly}
M.J. Hartmann, F.G.S.L. Brand{\~a}o, and M.B. Plenio, Nat. Phys. {\bf 2}, 849 (2006).

\bibitem{greentree2006quantum}
A.D. Greentree, C.~Tahan, J.H. Cole, and L.C.L. Hollenberg, Nat. Phys. {\bf 2}, 856 (2006).

\bibitem{DWave} 
M. W. Johnson, et al., Nature {\bf 473}, 194 (2011).

\bibitem{Houck2012} 
A. A. Houck, H. E. T\"ureci, and J. Koch, Nature Phys. {\bf 8}, 292 (2012).

\bibitem{Koch2013}
S. Schmidt and J. Koch, Ann. Phys. (Berlin) {\bf 525},  395 (2013).

\bibitem{Steffen2013} 
J. M. Chow, J. M. Gambetta, E. Magesan, S. J. Srinivasan, A. W. Cross, D. W. Abraham, N. A. Masluk, B. R. Johnson, C. A. Ryan, and M. Steffen, Nat. Commun. {\bf 5}, 4015 (2014).

\bibitem{Koch2007}
J. Koch, T. M. Yu, J. Gambetta, A. A. Houck, D. I. Schuster, J. Majer, A. Blais, M. H. Devoret, S. M. Girvin, and R. J. Schoelkopf, Phys. Rev. A {\bf 76}, 042319 (2007).

\bibitem{Blais2012}
J. Bourassa, F. Beaudoin, J. M. Gambetta, and A. Blais, Phys. Rev. A {\bf 86}, 013814 (2012).










\bibitem{Leo2010}
L. DiCarlo, M. D. Reed, L. Sun, B. R. Johnson, J. M. Chow, J. M. Gambetta, L. Frunzio, S. M. Girvin, M. H. Devoret, and R. J. Schoelkopf, Nature {\bf 467}, 574 (2010).

\bibitem{Ballester2012}
D. Ballester, G. Romero, J. J. Garc\'{i}a-Ripoll, F. Deppe, and E. Solano, Phys. Rev. X {\bf 2}, 021007 (2012).

\bibitem{Blais2011}
J. M. Gambetta, A. A. Houck, and A. Blais, Phys. Rev. Lett. {\bf 106}, 030502 (2011).

\end{thebibliography}
\end{document}